# Chloride-induced corrosion of steel in concrete – insights from bimodal neutron and X-ray microtomography combined with ex-situ microscopy

Ueli M. Angst [a*], Emanuele Rossi [a], Carolina Boschmann Käthler [a,b], David Mannes [c], Pavel Trtik [c], Bernhard Elsener [a], Zhou Zhou [d], Markus Strobl [c,e]

[a] Institute for Building Materials, ETH Zurich, Zurich, Switzerland
[b] Hagerbach Test Gallery Ltd., VSH, Flums, Switzerland
[c] Laboratory for Neutron Scattering and Imaging (LNS), Paul Scherrer Institut, Villigen, Switzerland
[d] Department NPM2/RST, Faculty of Applied Sciences, Delft University of Technology, Delft, The Netherlands
[e] Niels Bohr Institute, University of Copenhagen, Copenhagen, Denmark

* corresponding author: uangst@ethz.ch

## Abstract

The steel-concrete interface (SCI) is known to play a major role in corrosion of steel in concrete, but a fundamental understanding is still lacking. One reason is that concrete's opacity complicates the study of internal processes. Here, we report on the application of bimodal X-ray and neutron microtomography as in-situ imaging techniques to elucidate the mechanism of steel corrosion in concrete. The study demonstrates that the segmentation of the specimen components of relevance – steel, cementitious matrix, aggregates, voids, corrosion products – obtained through bimodal X-ray and neutron imaging is more reliable than that based on the results of each of the two techniques separately. Further, we suggest the combination of tomographic in-situ imaging with ex-situ SEM analysis of targeted sections, selected on the basis of the segmented tomograms.

These in-situ and ex-situ characterization techniques were applied to study localized corrosion in a very early stage, on reinforced concrete cores retrieved from a concrete bridge. A number of interesting observations were made. First, the acquired images revealed the formation of several corrosion sites close to each other. Second, the morphology of the corrosion pits was relatively shallow. Finally, only about half of the total 31 corrosion initiation spots were in close proximity to interfacial macroscopic air voids, and >90% of the more than 160 interfacial macroscopic air voids were free from corrosion. The findings have implications for the mechanistic understanding of corrosion of steel in concrete and suggest that multimodal in-situ imaging is a valuable technique for further related studies.

(248 words)

# 1. Introduction

Corrosion of steel in concrete is the most common degradation mechanism limiting the life of reinforced concrete structures [1], which causes high socio-economic costs, threatens personal safety, and increases the consumption of natural resources and energy [2-4]. It is thus important to fundamentally understand the mechanism of steel corrosion in reinforced concrete structures. However, the study of corrosion processes of steel in concrete is hampered by the fact that concrete is an opaque, nontransparent medium. This circumstance significantly narrows the array of tools available for in-situ studies of the corrosion initiation and propagation processes. While a range of electrochemical methods can be applied to monitor, generally non-destructively, the corrosion state of the steel, these techniques provide limited or no spatial information. For instance, the exact location where corrosion reactions occur on steel in concrete is difficult to assess with electrochemical techniques. The microscopic and macroscopic material heterogeneities at the steel-concrete interface (SCI) are another key aspect influencing the mechanism of corrosion in reinforced concrete, which however can hardly be characterized. Such spatial information about the location of corrosion reactions and their surrounding concrete features is fundamental for the study of the mechanism of localized corrosion (e.g. chloride-indued corrosion) of steel in concrete [5, 6].

Thus, to obtain spatial information related to corrosion phenomena in concrete, techniques for ex-situ analyses are generally used. These techniques, however, involve destructive sample preparation such as cutting and polishing sections for optical and scanning electron microscopy [7, 8]. The microscopic studies are often combined with other characterization techniques, particularly spectroscopic techniques including Raman, energy-dispersive X-ray, etc. The sample preparation steps required for ex-situ analyses are known to potentially disturb the conditions within the concrete, e.g. introducing cracks during cutting (mechanical damage) and drying (shrinkage), or due to the introduction of heat or water during cutting and polishing [9-11]. Moreover, it may be difficult to interpret the observations that can be made by the inspection of sections or split samples. This is because the two-dimensional information of sectioned near-spherical concrete components such as voids or aggregates can be misleading both with respect to their size and location. Another weakness of ex-situ analyses is that reference state information prior to the occurrence of corrosion is not accessible. In the generally post-corrosion ex-situ observations, precipitated corrosion products may mask voids or other features that may have been present before corrosion started [5]. These limitations can lead to misinterpretations about the role of different steel-concrete interfacial features in the corrosion mechanism.

A technique increasingly used to overcome the limitations of ex-situ analyses is X-ray computed microtomography (X-ray microCT). The advantages related to its application are that X-ray microCT requires almost no sample preparation and that monitoring over time is possible. This technique is powerful in imaging the steel, and thus providing information about related changes, such as the growth of corrosion sites [12, 13]. X-ray microCT has been successfully used to analyze the consequences of corrosion of steel reinforcement in concrete, such as build-up of corrosion products at the SCI and micro-cracking development [14, 15]. However, a weakness of X-ray microCT is its limited ability to characterize the concrete matrix in close vicinity of the steel, thus in the interfacial region. This is because, firstly, the cement paste and concrete aggregates have very similar X-ray attenuation coefficient. Furthermore, the steel-concrete interfacial zone is prone to X-ray metal artifacts due to, for instance, the pronounced mismatch in X-ray photoelectric absorption between the concrete and the steel, and due to X-ray beam hardening effects of polychromatic X-ray beams [11, 16-18]. These effects may distort the acquired signals and the reconstructed images exactly in the zone of interest for the study of corrosion processes of steel in concrete. As a result, the related unphysical biases in grey values in the obtained images present major difficulties for further image analysis and interpretation, such as segmentation. This is an important limitation because concrete-related properties of the SCI, such as the presence of voids and their moisture state, play a major role in corrosion [5, 19, 20].

To reliably characterize not only the steel part of the SCI, but also the adjacent cementitious matrix, including features such as pores and precipitated corrosion products, we here show results from a multimodal in-situ imaging approach, combining neutron and X-ray tomographic imaging modalities. The main advantage of fusing the data from these two imaging modalities is that this allows a more precise and reliable segmentation of all phases needed to elucidate the mechanism of localized corrosion of steel in concrete than segmentation based on only one of the two modalities. The relevant phases that need to be segmented include the steel, cement hydration products, aggregates, corrosion products, and voids. The more reliable segmentation as compared to single mode imaging (e.g. neutron or X-rays on their own) can be traced back to the fact that neutron and X-rays significantly differ in terms of their attenuation behaviour. Neutrons are indeed strongly attenuated by elements like hydrogen, but can better penetrate many metals [21]. Therefore, neutron imaging can be used to study water content and moisture transport in cementitious materials [22-24], but it has been rarely used to study corrosion of steel in concrete [25]. In contrast to neutrons, the X-ray attenuation correlates with the electron density (atomic number) of a material and, thus, it is significantly attenuated by metals like steel. The combination of both markedly different attenuation behaviors results in a complementary image of the specimens. A successful application of this multimodal technique to image reinforced concrete specimens undergoing corrosion has been reported by Robuschi et al. [26], which can be further discussed and developed to elucidate the corrosion mechanism of steel in concrete and the related consequences for engineering structures. The need of further micro-scale and over-time studies on the present topic comes from, among other reasons, the fact that in most of studies where X-ray CT was applied to investigate corrosion of steel in concrete, the experimental conditions were generally far from being realistic. For instance, some specimens analyzed in previous studies were cast *ad hoc* to optimize the X-ray image quality, embedding smooth steel bars with a diameter ranging between 0.4–3 mm [27, 28]. Other studies analyzed the corrosion progress of steel bars with a more practice-related diameter of 6–12 mm, but in most of the cases their corrosion was induced by galvanostatic polarization [29, 30], hence modifying the initiation and propagation mechanism that would occur under practice-related conditions. Furthermore, most of the specimens were cast with relatively small geometries and under laboratory conditions, likely resulting in specimens with characteristics not necessarily representative for engineering practice. Analysis of naturally corroded steel bars with practice-related dimensions and morphology and embedded in concrete have been reported in [15, 26]. Nevertheless, the specimens analyzed in these studies showed relatively late-stage corrosion damage, that is, there were significant amounts of corrosion products detected in the cementitious matrix and corrosion must have propagated in these samples for relative long times prior to the imaging.

Here, we report results from bimodal X-ray and neutron tomographic imaging on specimens retrieved from engineering structures, subsequently subjected to chloride-induced corrosion under controlled laboratory conditions, in order to obtain specimens with practice-related steel-concrete interfacial conditions and in an early corrosion-state, shortly after corrosion initiation. Moreover, the non-destructive, in-situ X-ray and neutron imaging data is compared to destructive, ex-situ microscopy. Based on the observations made, we discuss opportunities and limitations related to the used tomographic imaging modalities and draw conclusions related to the mechanism of corrosion-initiation of steel in concrete.

## 2.    Materials and methods

### *2.1.    Reinforced concrete samples and corrosion testing*

Four concrete cores were retrieved from the edge beam (top side) of a reinforced concrete bridge in Switzerland (labelled as B1-E1), the construction of which was completed in 2004 (Figure 1a). At the time of sampling, the age of the structure was ~15 years. The diameter of each concrete core was 150 mm, and it contained a ribbed reinforcing steel bar of diameter 10 mm (Figure 1b). The rebar was oriented horizontally in the structure (first reinforcing steel layer). No sample was corroding at the time of coring, as evidenced by potential measurements [31] on the structure prior to sampling. More information about the structure from which specimens have been collected are reported in Table 1.

*Table 1. Information about investigated structure, namely material properties and exposure conditions.*

| **Structure ID** | B1-E1 |
|---|---|
| Year of construction | 2004 |
| Structural element | Bridge edge beam |
| Core diameter (mm) | 150 |
| Steel bar diameter (mm) | 10 |
| Type of reinforcement | Carbon steel, presumably quenched and self-tempered |
| Orientation of reinforcement | Horizontal, 1$^{st}$ layer from the top |
| Concrete properties | Ordinary Portland cement; entrained air; w/c=0.45-0-5 |
| Height of structural element (mm) | Approx. 600 |
| Exposure conditions | XD3, chloride exposure from the top |
| Location of the structure | Swiss Alps |
| Height above sea level (m) | 1100 |

The concrete samples were transferred to the laboratory (ETH Zurich, Switzerland) and underwent a number of sample preparation steps, including: 1) reducing the concrete cover from initially 80 mm to 15 mm by water-cooled diamond saw cutting, 2) establishing an electrical cable connection to the rebar, 3) protecting the end parts of the rebars from crevice corrosion by means of a protocol described in Ref. [32], and 4) coating the lateral sides as well as part of the exposed side of the core with an epoxy resin to ensure that only 60 mm of the concrete surface along the rebar was exposed to chlorides in the subsequent corrosion test.

To induce corrosion of the embedded reinforcement, the reinforced concrete samples were placed in a 10% sodium chloride solution to allow for chlorides to diffuse into the concrete and to trigger corrosion of the steel bar (Figure 1c). The electrochemical potential of the embedded steel bar was continuously monitored with the help of a reference electrode and a datalogger (multimeter with input impedance >1 MOhm), until a drop of the potential indicated initiation of corrosion, which occurred after around 2 years from the start of the exposure of the cores to the sodium chloride solution (Supplementary Figure S1). A more detailed description of the sample preparation protocol and the corrosion test can be found in Ref. [32].

## 2.2. In-situ and ex-situ characterization

### 2.2.1. X-ray and neutron image acquisition/computed tomography

After corrosion initiation, a core of 25 mm diameter was extracted from the tested specimen, such that the entire 150 mm long rebar was centrally located in the 25 mm core (Figure 1d). A total of four such cores were obtained, labelled as B1-E1-H-1 to B1-E1-H-4. The cores were transferred to the facilities for neutron and X-ray microtomography at the TU Delft (X-Ray microCT: Civil Engineering and Geosciences, neutron microCT: Reactor Institute Delft) as soon as corrosion initiated, so that the mechanism of corrosion of steel in concrete could be studied in a very early stage (shortly upon chloride-induced corrosion onset) and in-situ, that is in a largely undisturbed state (Figure 1e).

X-ray microCT images were acquired through a Phoenix Nanotom (Waygate Technologies, GE, Germany) with a transmission acceleration voltage of 150 kV and an exposure time of around 85 minutes. Prior to acquiring the X-ray images, calibration of the projections with dark and bright field images was conducted. The X-ray CT scan was composed of 2303 slices in the height direction (Z-axis) and 2283 slices in width and depth directions (X- and Y-axis respectively). The reconstruction of X-ray images was performed through a Phoenix Datos|x software, and correction of ring, spot and beam hardening artifacts was conducted. The voxel size of the reconstructed images had a size of 25 μm and it contained 16 bits.

The neutron images were acquired at the FISH neutron imaging station at HOR (Delft, the Netherlands) with 626 projections over 360 degrees. The exposure time was 65 seconds per projection, with the chip cooled to –40°C for noise reduction. The detector system is based on neutron scintillators coupled via mirror and lens system to a CCD/sCMOS camera in a light tight box. The neutron scintillator consisted of a $ZnS/^{6}LiF$ with a thickness of 200 μm from RC TRITEC (RC TRITEC AB, Switzerland). As camera system, an Andor NEO-5.5-CL3 with 2560 by 2160 pixels was used. The field of view was 100 mm by 50 mm and the corresponding voxel size was 49 μm. More information about the FISH neutron imaging station can be found elsewhere [33]. The scattering corrections have been performed using an established black bodies methodology [34, 35].

The reconstructed X-ray and neutron CT images were registered (i.e., aligned) in the CT analysis software VG studio max 3.3 (Volume Graphics GmbH, Heidelberg, Germany). In the volume data, sample features such as *voids* (air voids, plastic settlement zone), *corrosion products*, and *steel* show considerable differences. Thanks to the complementary nature of the neutron and X-ray image data, the grey scale values (GSV) representative of the attenuation coefficients of the specimens' components vary considerably between the two data sets. Hence, image segmentation was carried out using the data set where the feature of interest was more pronounced. The steel was segmented on the basis of the grey value (attenuation) histogram of the X-ray-images, while the corrosion products were segmented with the grey value histogram of the neutron-images, using the surface determination tool respectively the defect analysis tool of VG studio max. The two resulting regions of interest representing the investigated features, have additionally been locally corrected with the region growing tool, which allows image segmentation by choosing connected neighboring pixels based on the initial seed points. The segmentation of air voids was conducted manually with the region growing tool in the neutron-images and subsequently validated comparing the neutron-based segmented areas to the respective X-rays-based areas. Furter information on the segmentation using combined information from X-ray and neutron attenuation can be found in the Supplementary Information (section A.2).

### 2.2.2. Sample preparation and Scanning Electron Microscopy (SEM) analysis

After tomographic in-situ imaging (Figure 1e), the samples were transferred back to ETH Zurich, where selected sections were studied by conventional microscopy (Figure 1f). These sections were selected based on the results of the segmentation of the bimodal microtomography to study sections with features of particular interest such as locations where corrosion initiated or where air voids were found close to the SCI. With the knowledge about the coordinates of these features within the sample, sections were cut with a water-cooled diamond saw. After sawing, the surface of the sections was firstly ground and then polished to further optimize the location of the sampled section before imaging with a scanning electron microscope (SEM). To minimize damage related to the sample preparation, a protocol described and validated elsewhere [11] was carefully followed. Microscopical analysis were performed through a FEI Quanta 600 environmental SEM with a Backscattered Electrons detector (BSE). Micrographs were acquired with an acceleration voltage of 20 kV and a magnification of x85 and x300.

An overview of the image acquisition settings used in the X-ray and neutron microtomography and SEM analysis is reported in Table 2, while a schematic representation of the whole experimental procedure previously described is visible in Figure 1.

Table 2. Selected details about the used in-situ neutron and X-ray tomography acquisition techniques, and the ex-situ SEM microscopy technique.

|  | Image acquisition technique | | |
| --- | --- | --- | --- |
|  | X-ray tomography (in-situ) | Neutron tomography (in-situ) | SEM microscopy (ex-situ) |
| Pixel/voxel size | 25 μm | 49 μm | submicron |
| Total exposure time [min] | 85 | 680 | N.A. |
| Exposure time (neutron) [s] | N.A. | 65 | N.A. |
| Acceleration voltage [kV] | 150 | N.A. | 20 |
| Field of view (neutron) [mm] | N.A. | 100 x 50 | Approx. 2 x 2 |

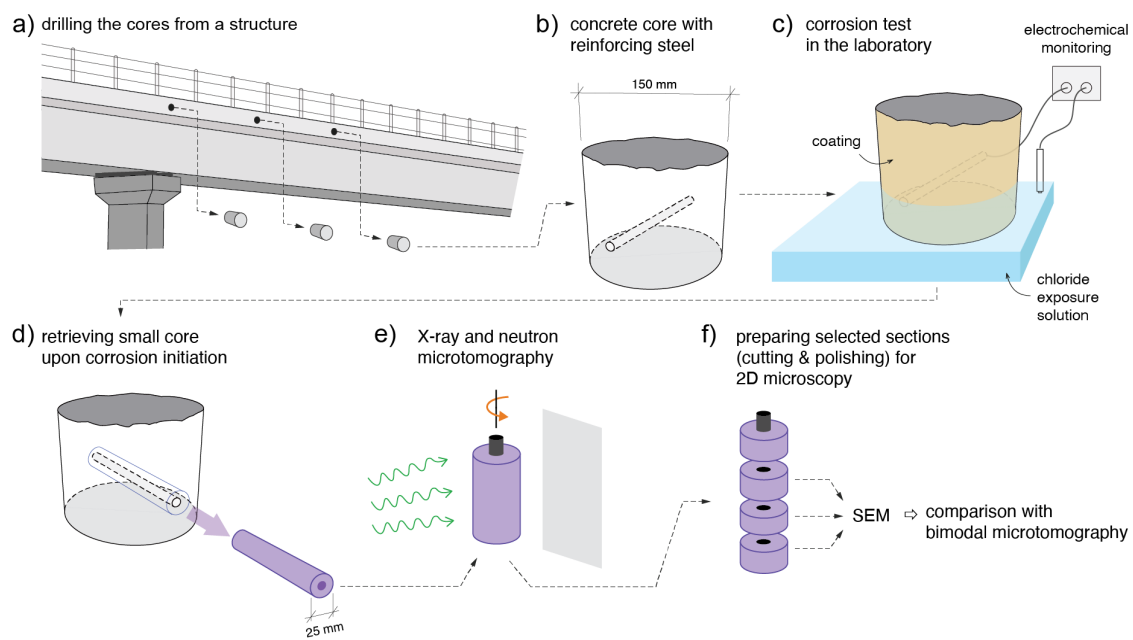

Figure 1. Sequence of experiments from retrieving samples from a reinforced concrete bridge (a,b) over corrosion testing in the laboratory (b) to the in-situ study of the samples in an early stage of corrosion by means of X-ray and neutron microtomography (d,e) and comparison with classical ex-situ microscopy (f).

# 3. Results and discussion

## 3.1. In-situ and ex-situ characterization and imaging

An example of reconstructed vertical and horizontal sections of both X-ray and neutron imaging of the reinforced concrete sample is shown in Figure 2. The reinforcing steel bar is clearly visible in the X-ray imaging mode due to the pronounced X-ray attenuation behavior of steel. Furthermore, voids (black) and cement paste/aggregates (dark grey) can be distinguished in this imaging mode, but further distinctions between the solid phases constituting the concrete are difficult to be performed due to the noise and shadows consequent to the X-rays exposure of the specimen. On the other hand, the neutron imaging mode allows the distinction of hydrated cement phases from voids and aggregates. Additionally, a major difference in the two modes is the creation of artifacts in X-ray imaging near the rebar. This is clearly visible as dark areas due to the presence of the ribbed rebar surface, while the neutron imaging mode is largely free from such artifacts.

It should be noted that the visual distinction of the different features based on grey values and shape may in Figure 2 as shown here, namely in the form of a converted and compressed image format, possibly even printed on paper, not work as well as this was possible when the original images were viewed on a high-quality computer screen and with specialized software (VG studio).

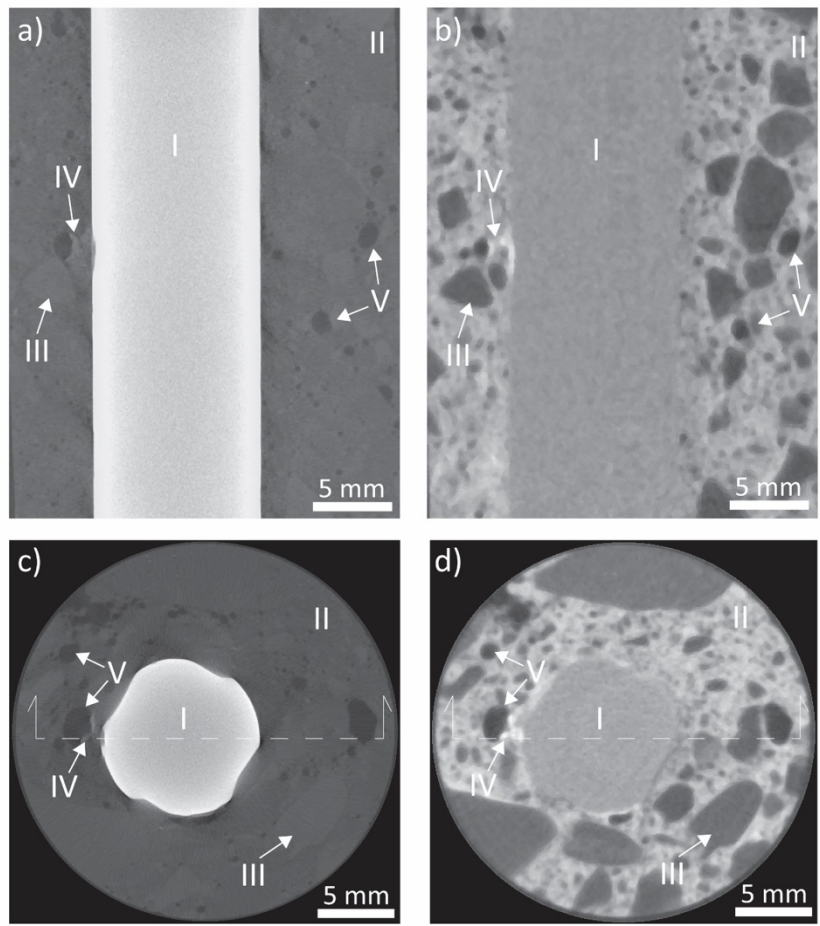

*Figure 2. Comparison of reconstructed sections of the two imaging modes X-ray (a,c) and neutrons (b,d), illustrating the pronounced differences in attenuation (grey level) between the two modes for different phases, such as the rebar (I), cementitious phases (II), aggregates (III), corrosion products (IV), and voids (V). The dotted lines visible in the horizontal slices of X-rays and neutron images, respectively, indicate the cross-section lines of the respective vertical images. Image analysis and detection of components were conducted on high-resolution data, of which image quality may differ from the one herein visible.*

Figure 3 shows an example of one inspected horizontal section where conventional ex-situ scanning electron microscopy was compared to the corresponding section as reconstructed from the X-ray and neutron tomography, respectively. The SEM micrograph shown here was composed by image stitching (from recording numerous high-resolution images along the SCI region). The SEM image in Figure 3 was adjusted to ensure identical size and orientation of the rebar and the features in the concrete matrix between all three imaging modes. It becomes apparent that the resolution of the images differs considerably. SEM yields the highest resolution of the different imaging modes, clearly in the submicron range. The pixel size of the microtomography was 25 and 49 μm for X-ray and neutron imaging modalities, respectively. By directly comparing the reference images acquired through X-ray and neutron microtomography and the respective SEM micrograph representative of the same specimen portion, a number of observations can be made based on the geometry of the specimen herein analyzed as well as on the specific image acquisition properties:

- Both X-ray and neutron imaging allow the identification of the steel reinforcement, the surrounding concrete matrix, and the voids inside the specimen. Neutron imaging allows distinguishing between different constituents of the concrete (e.g., aggregates) better than X-rays.

- X-ray shows considerable artefacts surrounding the rebar, where both darker and brighter regions are clearly apparent, which are absent in the SEM and neutron images. These artifacts are more pronounced than those observed in similar studies where cylindrical steel bars were used [15, 26, 36, 37] because of the ribbed geometry of the reinforcement embedded in the specimen herein investigated. Nevertheless, the availability of neutron images and SEM micrographs allows for a precise distinction of the constituents of the SCI. For specimens with similar geometry and characteristics, the calibration of X-ray images with complementary techniques is therefore advised.

- Through both X-ray and neutron imaging, macro-air voids (overall larger than around 80–200 μm) and corrosion product build-up at the SCI can be detected. However, the volume of corrosion products may be mis-estimated due to the interfacial artifacts in the case of X-ray imaging and by the similar gray scale value to that of hydrogen-rich components (e.g., cement hydration products) in the case of neutron imaging. Corrosion attacks at the reinforcement can also be reliably detected, especially in X-rays images. The extend at which this detection can occur mainly depends on the resolution (i.e., voxel size) of the image stack. In the present case, the smallest corrosion attacks that could be detected were around 50 μm deep.

- The detection of interfacial macro air voids with a diameter smaller than around 80–200 μm through X-ray and neutron microtomography was hampered. At the right and top-right side of the reinforcement, a few interfacial voids are indeed visible in the SEM micrograph but not according to the respective X-ray and neutron image. In the case of X-rays, the detection of interfacial voids was hindered by image noise and shadows at the SCI. In the case of neutrons, the hampering was caused by the need of a voxel size lower than the 49 μm herein obtained as well as by the ring artifacts coming from both the neutron image acquisition procedure and the specimen geometry. As further apparent from the SEM images reported in Figure 4, these air voids are often separated from the steel reinforcement by a thin (10-20 μm thick) cement paste layer, which was also not visible from both the X-ray and neutron images.

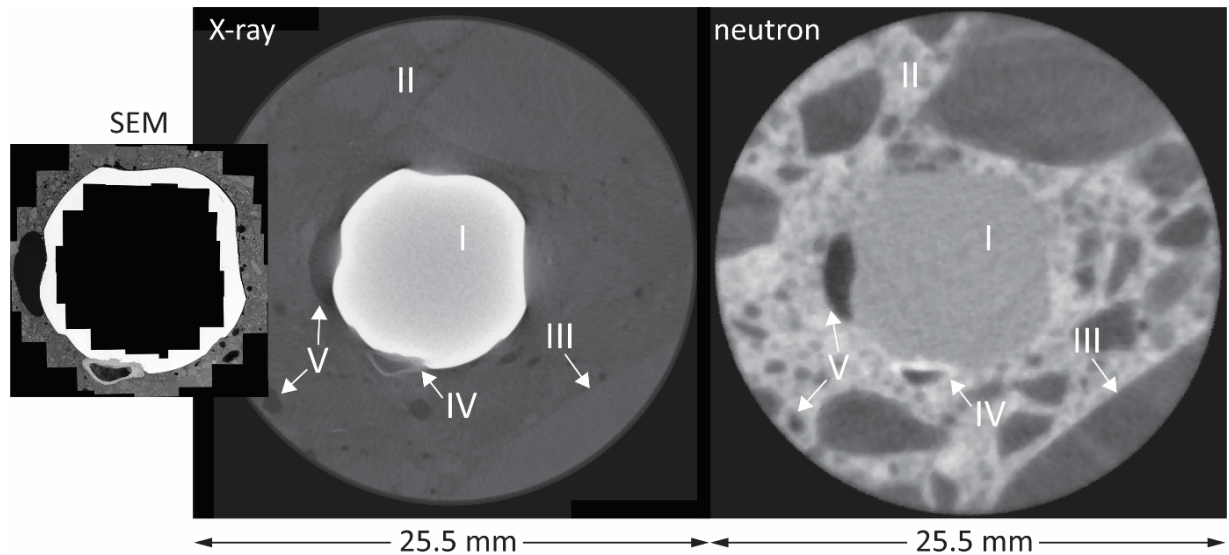

*Figure 3. Comparison (at identical size and orientation) of cut and polished section viewed in SEM (stitched image) with the corresponding reconstructed sections from X-ray and neutron imaging modalities, illustrating the different specimen phases such as the rebar (I), cementitious phases (II), aggregates (III), corrosion products (IV), and voids (V). Note that through SEM only the steel-concrete interfacial region was imaged, while in both X-rays and neutron images the concrete matrix of the core is fully visible. To better observe the characteristics of the SCI as well as the occurrence of corrosion pits, the SEM micrograph with higher magnification is also visible in Figure 8a. Note that the orientation of the reinforcement in the images may be different from the actual orientation of the steel when it was embedded in concrete, of which no information could be obtained.*

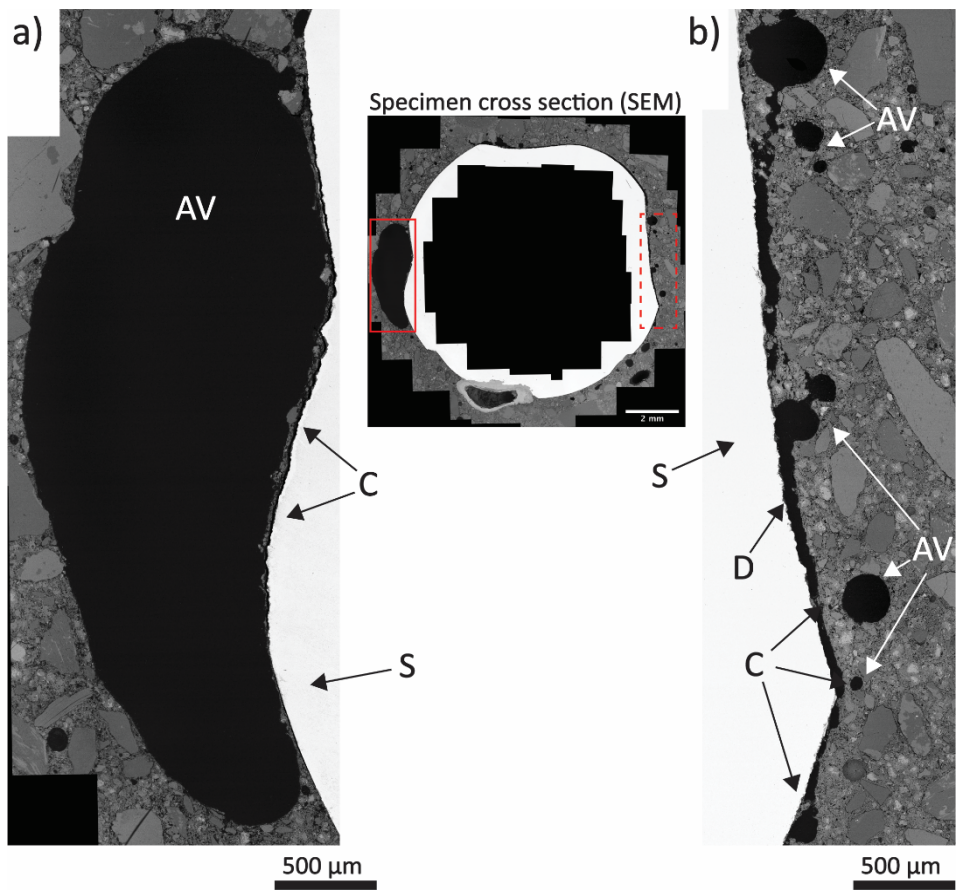

*Figure 4. SEM image of the specimen cross section (center) with areas where interfacial air voids are found at higher magnification, showing the different phases (S=steel; C=cement paste; AV=air void; D=debonding between steel and concrete; the solid and dotted red boxes highlight the portions reported at higher magnification in a) and b), respectively);*

*a) interfacial macro-air void, separated from the steel by a thin layer of cement paste of around 20 µm thickness; b) multiple small interfacial voids separated from the steel by a layer of cement paste with variable thickness.*

### 3.2. Image analysis and implications for the design of experiments

The identification of the different phases of a specimen through image thresholding can be a challenging process, especially when analyzing reinforced concrete specimens, of which some constituents have very similar attenuation coefficients when exposed to either X-rays or neutrons (i.e., cement paste and aggregates in the former case, hydrogen-rich components in the latter case). The potential artifacts related to the image acquisition procedure and specimen layout (e.g., ring noise, beam hardening effect) can furthermore lead to misinterpretation of the actual state and size of the specimen components. One strategy that can potentially reduce the risk of misinterpreting the internal state of reinforced concrete elements is to design specimens *ad hoc* to improve the image acquisition. This strategy would imply, among other aspects, locating the steel reinforcement centrally in the concrete core, using cylindrical bars with limited diameter or with a hollow geometry to reduce the X-ray beam hardening effects, and keep the specimens in dry conditions to increase the sharpness and contrast of the specimen constituents [14, 17]. While these features would improve the acquisition and reconstruction of the microtomography images and the later processing operations, it is obvious that most of these conditions cannot be met when collecting a sample from an existing structure, as the one analyzed in the present study, and when targeting realistic conditions in general.

Besides the artifacts that may result from the exposure of reinforced concrete specimens to either X-rays or neutrons, an important parameter to be considered when performing microtomography is the resulting pixel (or voxel) size of the acquired images. The smaller the voxel size, the higher the level of details of the image but, on the contrary, the smaller the scanned portion. Therefore, the resulting voxel size is a compromise between these parameters. Reducing the specimen size as much as possible would improve the level of detail of the images, but the results of small specimens might be not fully representative of conditions encountered in engineering structures. In any case, some information related to the internal composition of the reinforced concrete specimen may not be gathered because of the imaging artifacts and the dimension range of smaller characteristics. In the specimen analyzed in the present study, this is the case for interfacial voids smaller than around 80–200 µm and the 10–20 µm thick cement paste layer that often separates interfacial voids from the steel surface, which could be detected and observed thanks to the SEM micrographs visible in Figure 3-4. To have a more precise overview of the SCI and its small characteristics, conducting SEM analysis to validate what can be observed from X-ray and neutron images is highly recommended, especially when the aim of the study involves the micro-structural and microscopic characterization of the interfacial zone.

To improve the image segmentation and, therefore, the identification of different components of the specimens, a bivariate histogram composed of the X-ray and neutron attenuation coefficients can be used (Figure 5). By coupling the X-ray and neutron attenuation coefficients in a bivariate histogram, regions of pixels (or voxels) can be segmented based on the behavior that the representative elements have when exposed to both X-rays and neutrons. Given the complementarity of the information that the two techniques provide when applied to reinforced concrete specimens, a more reliable segmentation can be potentially obtained through this methodology [38].

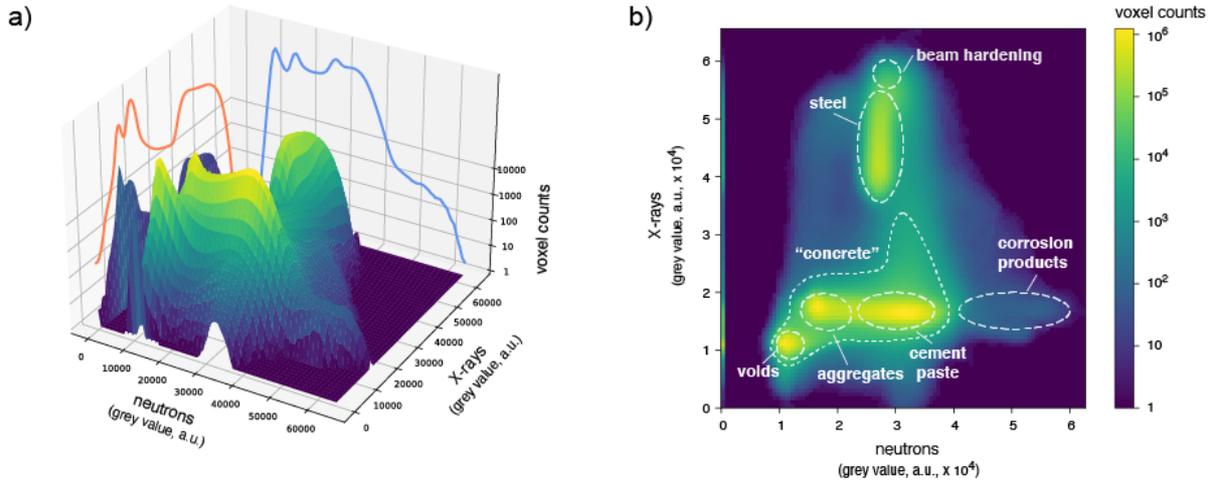

*Figure 5. Bivariate histogram displaying the number of voxels associated with certain X-ray and neutron attenuation intensities of the whole image stack (representative of the whole specimen); a) three-dimensional visualization, b) two-dimensional plot with indicated interpretations of the combined X-ray and neutron attenuation behavior in terms of features encountered.*

In Figure 5 a bivariate histogram of the number of voxels associated with certain X-ray and neutron attenuation coefficients is visible for the whole image stack of a here imaged specimen. A three-dimensional visualization of the two X-ray and neutron attenuation coefficient histograms is visible in Figure 5a, while the respective two-dimensional plot is reported in Figure 5b. In this latter case, it is possible to identify areas that belong to certain components of the specimen based on the behavior observed when simultaneously exposed to X-rays and neutrons (e.g., the attenuation coefficient measured for each voxel) and, therefore, improve the reliability of the segmentation. Elements like air voids can be easily identified in the region of $1-1.3 \times 10^4$ grey value, given their small X-ray and neutron attenuation and their consequent darkest color in the collected images. Similar operations can be conducted to identify and quantify the steel reinforcement as well, which has the highest X-ray attenuation coefficient ($3.7-5.5 \times 10^4$ gray value) and, therefore, it can be easily segmented. Since steel has a similar neutron attenuation coefficient to that of some portions of the cement paste, a reliable segmentation of the reinforcement based on the neutron image stack would be a challenging process, highlighting the potential of this bimodal methodology. The same issue occurs when aggregates, cement paste, and corrosion products must be segmented in X-ray images. Given the slightly higher X-ray attenuation coefficient of corrosion products, a segmentation of them may be possible but likely sensitive to inaccuracies when conducting quantitative analysis. A way more complex process would be segmenting the cement paste and the aggregates separately, given their almost identical X-ray attenuation. However, the availability of the neutron images deeply facilitates the segmentation of these three components. Aggregates, cement paste, and corrosion products have three distinct neutron attenuation coefficients because of their different content of hydrogen, ranging between $1.5-2.2 \times 10^4$, $2.4-3.7 \times 10^4$ and $4.1-6.0 \times 10^4$ gray value, respectively. By combining all these information and the different response that the reinforced concrete elements have to X-rays and neutrons, a reliable and precise segmentation of each specimen component can therefore be obtained, as reported in the Supplementary Information (Figure S2-S3).

### 3.3. Mechanism of corrosion of steel in concrete

#### 3.3.1. Morphology of corrosion attack

The simultaneous application of X-ray and neutron microtomography results in a complete overview of the internal state and conditions of the scanned specimen, which can be further analyzed with respect to the corrosion mechanism of steel in concrete. To this end, a three-dimensional visualization of one selected interfacial zone with segmented features (e.g., the steel, the voids and the corrosion products) is shown in Figure 6. In Figure 6a, several corrosion "pits" (a more detailed discussion related to the adequacy of this term follows below) can be observed with the build-up of corrosion products in their vicinity. The same pits, namely P1, P2 and P3, are better visible in Figure 6b, where corrosion products are not displayed. The pits in the front side of the reinforcement (P1, P2) are without an air void in their immediate vicinity, whereas the pit more aside (P3) formed in the proximity of a small air void (at a distance of about 500 μm), as visible in Figure 6c. The influence of air voids will be discussed in section 3.3.2. Here, we will discuss features related to the size and geometry as well as the distance between individual corrosion attacks.

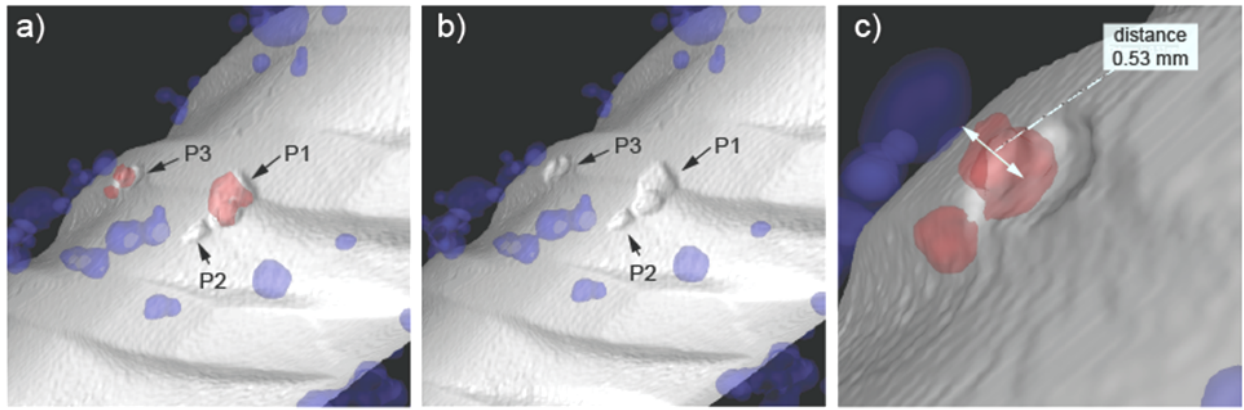

*Figure 6. Examples of segmented 3D renders illustrating interesting features at the steel-concrete interface (cementitious matrix and aggregates suppressed in the displaying mode for clarity): a) corrosion initiation pits in the rebar and presence of corrosion products (red) and macroscopic concrete voids (blue); b) same but without displaying the corrosion products, to better reveal the pits; c) illustration of a measurement of the distance between a macroscopic concrete void and a pit (higher magnification than a and b). Rebar diameter = 10 mm.*

Figure 6 shows relatively flat and shallow "pits" (in the range of 50-100 μm depth and 200-600 μm width and length). The here apparent morphology of the localized corrosion attack was representative of the numerous other pits observed in the other specimens. The ratio pit width:depth and pit width:length were generally clearly larger than 1. It may be questioned whether so shallow localized corrosion attacks may still be termed "pits", considering classical pitting corrosion theory. According to the theory underlying pitting corrosion of passive metals, the stability of newly formed pits is significantly affected by the geometry of the pits. When the local conditions to depassivate the steel are present (e.g., the necessary chloride concentration in relation to hydroxyl ions), metastable pits form and grow at a rate that depends on various factors, including metal composition, pit electrolyte concentration, and pit bottom potential [39]. While autocatalytically growing, the anodic and cathodic corrosion reactions separate spatially during pitting. The local pit environment becomes depleted in oxygen, which shifts most of the cathodic reaction to the exposed surface where there is higher availability of this reactant. A consequence of this classical pitting corrosion mechanism is the in-depth growth of a corrosion pit while the rest of the metal surface remains passive [40]. Transport processes between pit bottom (where ferrous ions are released) and the outside areas (where cathodic reaction occurs) are an important factor in pit growth, which is markedly affected by the geometry of the pit (depth, width) [41]. Moreover, the availability of negatively charged ions different from hydroxyl ions (such as chloride ions) is crucial in order to maintain stable pit growth, as the migration of hydroxyl ions into the pit may lead to repassivation [42, 43].

The observation of the shallow morphology of "pits" observed in this study has implications for the understanding of the mechanism of localized corrosion of steel in concrete. While the processes related to classical pitting, mentioned above, may still apply for localized corrosion in concrete, additional factors play a role and cause the growth to occur in lateral direction rather than primarily in depth. In this regard, authors have previously argued that the anodic and cathodic current distributions may cause the pit to grow outwards [44, 45]. However, this explanation based on differences in local current density may apply to pitting of a metal in an electrolyte similarly as to localized corrosion of steel in concrete, and thus, it may not fully explain the experimental observation of the shallow pit morphology. An additional factor may be related to the simultaneous release of both chloride and hydroxyl ions in the vicinity of the pit as a result of the local acidification caused by hydrolysis of iron ions released in the corrosion process. This acidification will to some extent be buffered by the presence of cementitious phases (e.g. Portlandite), that will then dissolve and release hydroxyl ions. At the same time, the acid-attack of cementitious phases (e.g. CSH or Friedel's salt) will release chloride ions. Depending on the rate and ratio of release of chloride and hydroxyl ions, the passive film next to the pit may be destabilized, which may promote outwards pit growth. Clearly, more research is needed to fully elucidate the mechanism.

The occurrence of corrosion "pits" close to each other as shown in Figure 6 may raise further interesting points of discussion. Admittedly, it is unclear from the available in-situ images, which were taken at one point in time, whether the different visible corrosion "pits" had formed in a sequence (the first one forming, then repassivating, followed by formation of another one, etc.) or whether they occurred simultaneously. Additional information on this question is available from the electrochemical monitoring of the corrosion process, especially the transition from passive state to active corrosion. The potential of the rebars was continuously monitored during chloride exposure. According to the experimental protocol, a decrease of at least 150 mV, maintained over at least 10 days, was used as the criterion to determine stable corrosion initiation [32]. Supplementary Figure S1 shows the potential vs. time during this corrosion initiation period for the different specimens. These data show a clear drop in potential (by 200-300 mV). Moreover, in 3 out of the 4 specimens, it was clearly apparent from the potential measurements that no repassivation events occurred over the following approx. 10 days; in one case (B1-E1-H-1), some slight potential variations were visible, however, no full repassivation was apparent. This electrochemical data thus suggests that a sequence of initiation of a pit, followed by repassivation of the same, subsequently followed by initiation of *another* pit, etc. was unlikely. Furthermore, since the observed pits have a similar size, we consider it unlikely that several pits occurred and always repassivated without this having been appreciable in the potential monitoring, until one more dominant pit had formed and caused the distinguished potential decrease that was generally observed. Thus, the systematic experimental observation of numerous corrosion "pits" close to each other likely is the result of *simultaneous* localized corrosion at different nearby locations.

Other authors have reported similar observations, that is, localized corrosion initiation sites in steel in concrete in close proximity. For instance, Wu et al. [46] observed the occurrence of corrosion attacks densely and patchily distributed along the steel reinforcement. Beck et al. also revealed the occurrence of "pits" in close proximity [13]. Angst and Elsener [6] inspected corroding bars by destructively splitting their specimens, and they observed the presence of several distinct corroding spots along the reinforcement within a surface a steel area of less than 1 $cm^2$ and in some cases at 500 μm distance.

These observations, and especially the indication that the localized corrosion sites likely occurred simultaneously rather than in sequence, may have important implications for the understanding of localized corrosion of steel in concrete. The formation and growth of a localized corrosion attack polarizes nearby locations of the passive film to more negative potentials. This is sometimes claimed to lead to some form of mutual protection, since it is known that the potential of (passive) steel impacts the chloride threshold [47]. Accordingly, some authors have suggested to consider this in service life modeling, namely to consider that a first anodic size would, to some degree, result in protection of the adjacent zones, with the hypothesized consequence that another pit is unlikely to occur in close proximity of the first one [48]. According to the current findings and to experimental studies that reported on the monitoring of corrosion pits formation over time, a different mechanism may occur. More precisely, the local characteristics at the SCI such as the presence of voids, the heterogeneity of both the steel surface and concrete composition, and the moisture conditions at the interfacial zone could be the dominant factor to cause corrosion pits growth at specific locations. Since all these parameters also influence the critical chloride content [5, 6], the observations reported in the present study suggest that a combination of these factors overrides the inhibiting effect of one corrosion attack polarizing the surrounding area. Thus, the presence of these characteristics may lower the chloride threshold at these locations, making them more prone to become active sites regardless of the local steel potential being shifted due to the occurrence of corrosion onset in the near vicinity. While at later stages of corrosion, where larger anodic zones have formed and may indeed lead to a certain degree of protection of adjacent zones, we here question the validity of this concept of mutual protection for early-stage corrosion.

It is worth mentioning that statistical evidence is needed to elucidate this mechanism and especially which are the leading factors that cause corrosion sites formation and growth, which is lacking in the present study. Nevertheless, these observations and the related points of discussion underline the need of more scientific understanding of the role that interfacial characteristics and defects have on the corrosion initiation and propagation mechanism.

### 3.3.2. Influence of interfacial microscopic and macroscopic voids

The role that interfacial voids have on the corrosion of steel reinforcement in concrete has been widely studied and discussed in the literature, as reviewed in [5]. Nevertheless, the complexity of the phenomenon as well as the interrelation between many influencing parameters are such that a comprehensive explanation of the mechanism has not been formulated yet. The three-dimensional data generated in the present study were evaluated with respect to the occurrence of corrosive attacks in relation to the characteristics of the SCI. More precisely, the formed corrosion sites identified along the steel bar as well as interfacial voids in the near proximity of the steel surface (i.e., in direct contact to the reinforcement or at a maximum distance of 100 μm) and with a minimum diameter of 80 μm were counted. Finally, the number of corrosion initiation spots in proximity of interfacial air voids was analyzed. An example of rendered steel volumes showing the occurrence of corrosion attacks at specific locations of the SCI is visible in Figure 7 for two specimens, while a summary of the analyses from the four specimens is reported in Table 3.

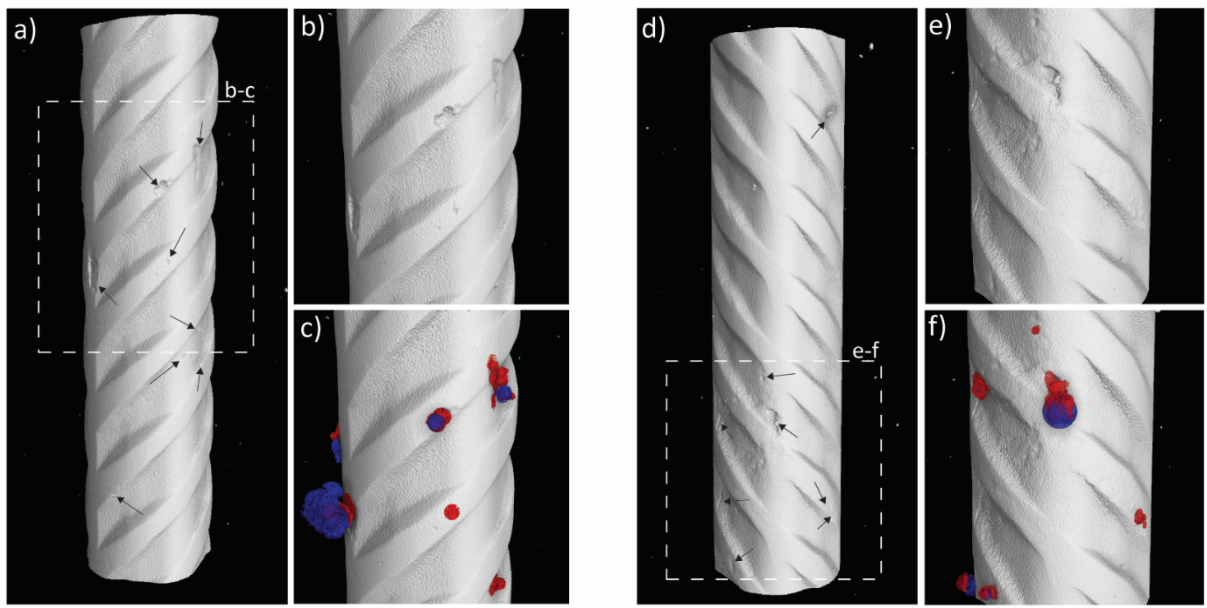

*Figure 7. Examples of segmented 3D renders illustrating interesting features at the steel-concrete interface (cementitious matrix and aggregates suppressed in the displaying mode for clarity): a) render of pristine steel of specimen B1-E1-H-2, indicating corrosion attacks with black arrows (the dotted square is the magnified portion of the specimen); b) higher magnification portion of the corrosion attacks of B1-E1-H-2; c) higher magnification portion of the corrosion attacks of B1-E1-H-2 showing corrosion products (red) and interfacial voids (blue) in the proximity of the corrosion attacks; d) render of pristine steel of B1-E1-H-3, indicating corrosion attacks with black arrows (the dotted square is the magnified portion of the specimen); e) higher magnification portion of the corrosion attacks of B1-E1-H-3; f) higher magnification portion of the corrosion attacks of B1-E1-H-3 showing corrosion products (red) and interfacial voids (blue) in the proximity of the corrosion attacks.*

Table 3. Information about the number of interfacial air voids and corrosion initiation sites observed in each specimen

| Sample-ID | Number of interfacial air voids^ without corrosion initiation | Total number of corrosion initiation spots | Number of corrosion initiation spots in proximity of interfacial air voids* |
|---|---|---|---|
| B1-E1-H-1 | 63 | 6 | 2 |
| B1-E1-H-2 | 54 | 11 | 5 |
| B1-E1-H-3 | 35 | 7 | 4 |
| B1-E1-H-4 | 14 | 7 | 4 |
| Total | 165 | 31 | 15 |

^ with a minimum diameter of 80 μm and a maximum distance from the steel reinforcement of 100 μm

* in direct contact to an air void or separated by a cement paste layer with a maximum thickness of 100 μm

From Figure 7 and Table 3, relevant information about the factors influencing the mechanism of corrosion initiation of steel in concrete can be obtained. First, in every specimen several initiation spots were observed, on average between 6 and 11 per imaged portion of each specimen. Second, a very large number of interfacial air voids was found (in total more than 150). Remember that these reinforced concrete specimens were taken from an engineering structure. It is important to highlight that out of the 31 identified corrosion spots, only about half of them occurred in the proximity of an interfacial void. It is also interesting to note that most interfacial air voids (>90%) were actually not characterized by corrosion attacks in their vicinity. These results indicate and confirm that the presence of air voids at the SCI do not necessarily create the most sensitive locations for corrosion to initiate per se, but that other intertwined factors play a fundamental role to create the conditions for corrosion to be triggered.

The results reported in Table 3 are in line with the findings related to the observations arising from the SEM observations reported below. In Figure 8 the SEM micrograph of a corroding section is visible, highlighting some regions of interest (ROIs) at higher magnification reported in Figure 9.

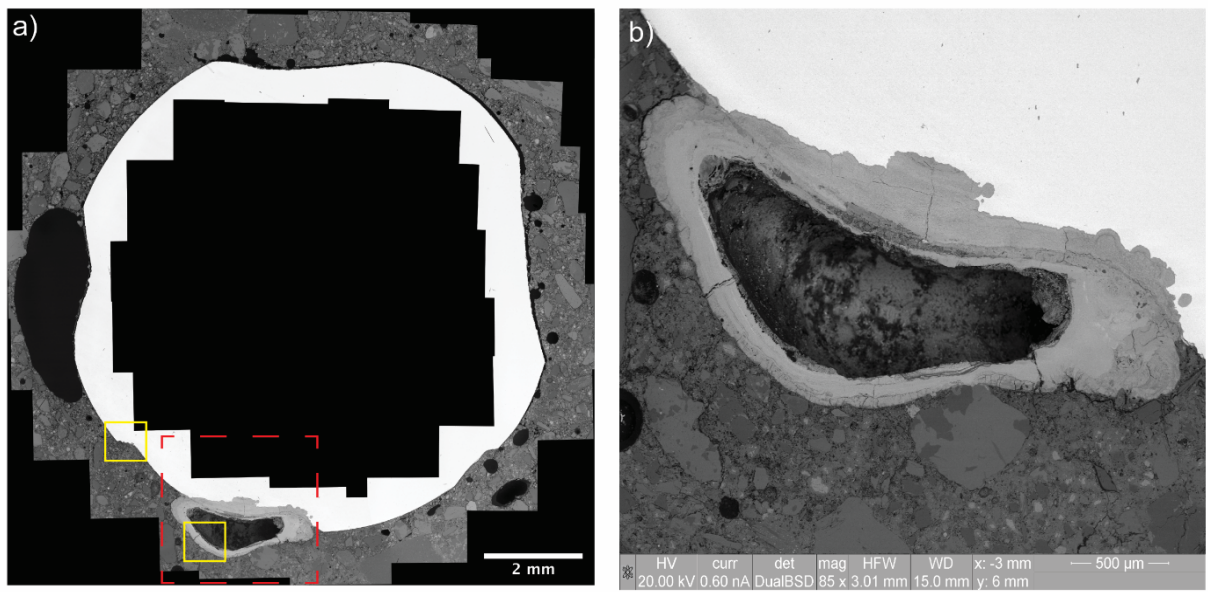

Figure 8. SEM micrograph of a corroding section (BSE detector, acquired at 20 kV and 85x magnification): a) overview of the corroding section made by stitching together the micrographs collected at the SCI, where the rebar is visible in white, the corrosion products in light gray, the concrete matrix in dark gray and the air voids in black (the red dotted-line box is representative of the zoomed-in area reported in Figure 8b, while the yellow boxes are areas shown at higher magnification in Figure 9a-b); b) Corrosion site in proximity to an interfacial air void, which is partially filled by corrosion products. Note that the orientation of the reinforcement herein reported may be different from the actual orientation of the steel when it was embedded in concrete, of which no information could be obtained.

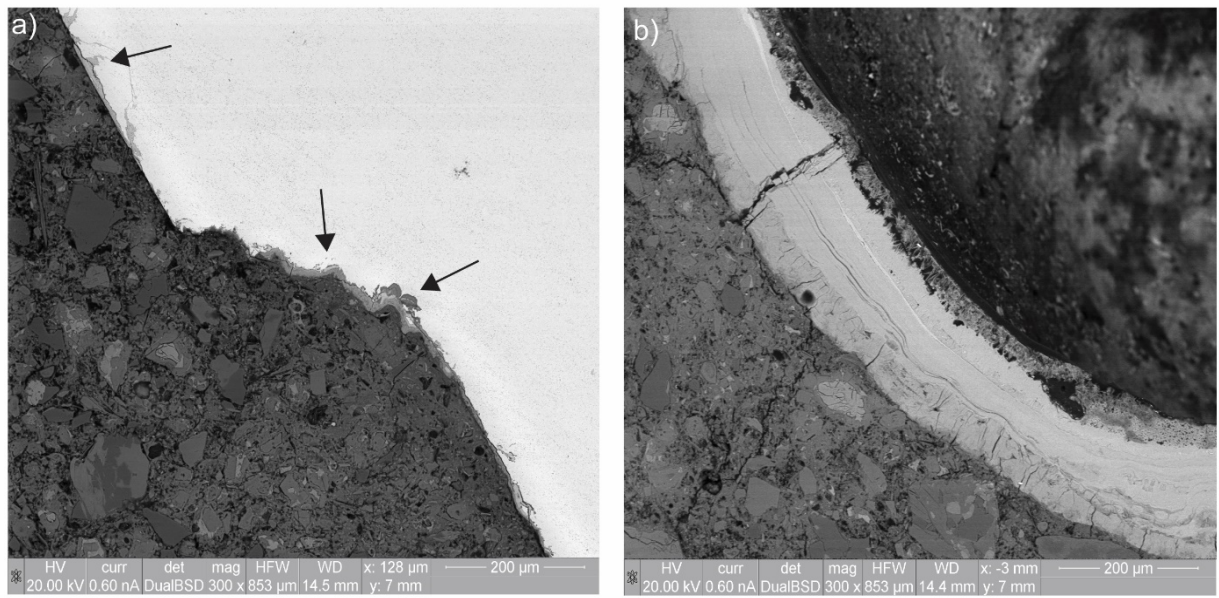

*Figure 9. SEM micrographs of corrosion pits at the SCI of a corroding rebar: a) small corrosion pits formed at the SCI without any macro defect present; b) layered morphology of corrosion products precipitated at the edges of an interfacial air void.*

As it is visible from Figure 8, different air voids are found at the SCI. Microscopic air voids with a diameter ranging between 50-200 µm are visible at the right and top-right side of the steel bar, often separated by a thin layer of cement paste (10-20 µm thick). A few small air voids are also visible at the bottom side of the steel. In the proximity of these interfacial microscopic voids, no corrosion sites can be observed. More interestingly, two macroscopic air voids are clearly visible at the left and bottom side of the bar, both significantly bigger than 1 mm. While the latter one is partially filled by corrosion products that precipitated into it from a corrosion pit in its vicinity, the former one is completely empty, and it is not in proximity of any corrosion pit. These observations are in line with the results reported in Table 3, according to which most microscopic and macroscopic air voids at the SCI did not per se create the most sensitive locations for corrosion to occur. Nevertheless, Figure 8 may suggest interesting points that need further analysis and experimental evidence to be clarified. Among others, an important reflection relates to the moisture conditions of the SCI. The growth of a corrosion site at one specific location and the lack of any corrosion attack where other interfacial voids are visible Figure 8 may suggest that the moisture conditions at the SCI may significantly vary around the steel cross section, regardless the exposure of the bigger scale specimen. Variations in moisture state of the concrete at the steel concrete interface upon capillary water uptake may occur depending on variability in pore structure [24].

The observations reported and discussed in this work highlight the need to study the corrosion mechanism more locally and at a higher level of detail, since a lot of open points are still to be clarified at a micro-scale level. Among others, these points include the inter-relation that appears between the locations at which corrosion attacks occur, the characteristics of the SCI, the presence of interfacial defects including their moisture state, and the speciation, transport, and precipitation of corrosion products consequent to corrosion initiation. With the present study, the potential of applying the described X-ray and neutron bimodal imaging method to elucidate the steel corrosion mechanism in concrete has been shown and elaborated. For the aspects that are still debated about the mechanism of corrosion of steel in concrete, the application of the X-ray and neutron bimodal imaging and SEM analysis can be cutting-edge, and therefore it will be implemented in future studies.

## 4. Conclusions

The present research reported on the applicability of a bimodal imaging technique based on simultaneous X-ray and neutron microtomography to study the corrosion mechanism of steel reinforcement in concrete. Limitations and advantages of this technique have been shown and discussed, also considering the addition of SEM analysis as a calibration tool to improve some features such as the segmentation of the different components of reinforced concrete specimens collected from existing structures. From the present study, the following major conclusions can be drawn:

- The segmentation of the specimen components obtained through bimodal X-ray and neutron imaging can be more reliable than that based on the results of each of the two techniques separately. Through a bivariate histogram, the image segmentation is based on both X-ray and neutron attenuation coefficients of each voxel, reducing the risk of misinterpretation of certain components by considering the different behavior of components when exposed to both imaging modalities.

- The in-situ X-ray and neutron microtomography techniques were combined with ex-situ SEM analysis of specific cross sections. While the SEM analysis confirmed the observations made with the in-situ tomographic techniques, the SEM imaging mode allowed studying various features in more detail and at higher resolution. We suggest the combination of in-situ tomographic X-ray and neutron imaging with ex-situ SEM analysis of targeted sections (selected on the basis of the in-situ studies) to maximize the information gain.

- From the SEM analysis, a significant amount of corrosion products precipitating in an interfacial macro air void was found, adjacent to the related corrosion pit. At the same time, other macro air voids with no adjacent pits nor precipitated corrosion products could be observed in the same cross section. This observation may suggest that the role of interfacial defects, which is believed to depend on many interrelated factors and mainly on the moisture conditions at the SCI, may significantly vary around the steel reinforcement cross section. Thus, the influence of the moisture conditions of the interfacial zone on the mechanism of steel corrosion must be locally and microscopically investigated.

- Preliminary analysis on the mechanism of corrosion sites formation have been conducted by studying localized areas at the SCI on the basis of the segmented tomograms. Three major observations were made based on this 3-dimensional information: i) Corrosion attacks were found very close to each other. This observation may suggest that for steel in concrete, corrosion onset does not necessarily and significantly protect the steel area in its proximity due to mutual polarization, but that the local susceptibility to corrosion initiation must be governed by factors related to the heterogeneity of the SCI. In this regard, it was in the four studied specimens found that ii) only about half (15) of the total (31) corrosion initiation spots were in close proximity to interfacial macroscopic air voids. Furthermore, there were more than 160 interfacial macroscopic air voids, and the vast majority (>90%) of them was free from corrosion. Finally, it was an interesting observation that iii) the corrosion attacks generally exhibited a shallow morphology. All these observations have implications for the understanding of the mechanism of localized corrosion of steel in concrete. More research to elucidate this mechanism is needed, and multimodal imaging, potentially in combination with ex-situ characterization, appears to be a valuable tool in this regard.

# Statements and declarations

**Competing Interests**

The authors declare that they have no known financial or personal conflicts of interest.

**Author contributions**

Ueli Angst, Carolina Boschmann, Pavel Trtik, David Mannes, and Markus Strobl conceived the overall study; all authors contributed to the study design. Experimental work was performed by Carolina Boschmann and Emanuele Rossi, and supported by Zhou Zhou, Pavel Trtik and David Mannes. Carolina Boschmann, Ueli Angst, Bernhard Elsener, and Emanuele Rossi performed the analysis and interpretation of the results in the context of corrosion. Ueli Angst wrote the main draft of the manuscript, to which all authors contributed. All authors read and approved the final manuscript.

**Acknowledgements**

The authors acknowledge the financial support from the Swiss National Science Foundation (projects no. PP00P2_163675 and PP00P2_194812) and the Swiss Federal Roads Office (project no. ASTRA AGB 2018/008). Furthermore, the authors are grateful to the Tiefbauamt Graubünden, Switzerland, for providing access to the reinforced concrete structure for sampling. The support by Dr. Nicolas Ruffray (ETH Zurich, Switzerland) with the acquisition of the scanning electron microscopy images and the support by Jeroen Plomp (Delft University of Technology, The Netherlands) with the coordination of the tomographic measurements are greatly acknowledged.

# Supplementary Information (SI) for manuscript

# Chloride-induced corrosion of steel in concrete – insights from bimodal neutron and X-ray microtomography combined with ex-situ microscopy

Ueli M. Angst [a]*, Emanuele Rossi [a], Carolina Boschmann Käthler [a,b], David Mannes [c], Pavel Trtik [c], Bernhard Elsener [a], Zhou Zhou [d], Markus Strobl [c,e]

[a] Institute for Building Materials, ETH Zurich, Zurich, Switzerland
[b] Hagerbach Test Gallery Ltd., VSH, Flums, Switzerland
[c] Laboratory for Neutron Scattering and Imaging (LNS), Paul Scherrer Institut, Villigen, Switzerland
[d] Department NPM2/RST, Faculty of Applied Sciences, Delft University of Technology, Delft, The Netherlands
[e] Niels Bohr Institute, University of Copenhagen, Copenhagen, Denmark

* corresponding author: uangst@ethz.ch

# A. Supplementary Information

## A.1. Potential measurements of reinforced concrete specimens

The potential measurements to electrochemically monitor the corrosion onset of reinforced concrete cores are reported in Figure S1.

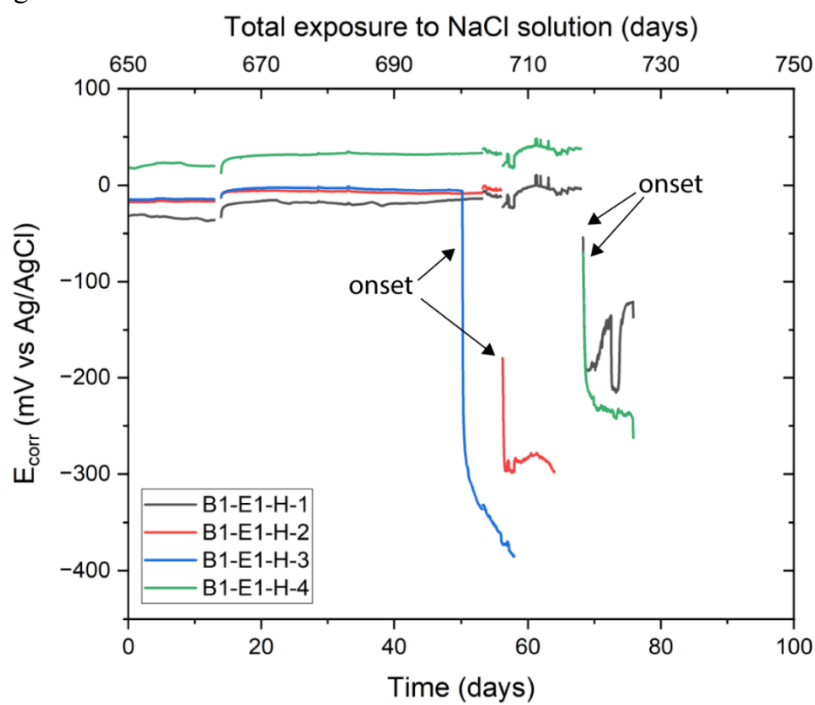

*Figure S1. Potential measurements of reinforced concrete specimens over time (vs. Ag/AgCl/KCl$_{sat}$), after which coring and X/n computed tomography scanning was conducted. Only the last measurements before corrosion onset are reported (the total duration of the exposure to NaCl solution is visible from the top x-axis)*

## A.2. Comparison between segmentation procedures based on singular or bivariate histograms

An example of the results of the segmentation procedure based on the images' gray-scale value (GSV) and bivariate histograms is visible in Figure S2-S3. For this example, a 2D cross-section image of the reinforced concrete specimen has been considered. Segmentation of both X-rays and neutron scans of the same cross sections has been performed by thresholding the respective GSV histograms (Figure S2), and the segmentation results are visible in Figure S3. Segmentation has been also performed using the bivariate histogram of the same cross-section (previously reported in the core manuscript, Figure 5). A K-Means clustering algorithm was used to differentiate the specimen's components as a basis for further segmentation. The pixel clusters related to aggregates, cement paste, voids, and background could be identified. On the other hand, the beam hardening effect present at the steel-concrete interface (SCI) made the segmentation between steel and corrosion products less sharp. To overtake this limitation, the steel region of the image was segmented via the X-ray GSV histogram, while the corrosion products were segmented via the neutron GSV histogram. The absence of any beam hardening effects would potentially make this procedure completely automatic. Since in the present paper no quantitative analysis was performed, the final segmentation allowed to gather the necessary information for the purpose of the study (e.g., measuring the corrosion attacks' dimensions, observing the interfacial characteristics where they occurred, counting and measuring the dimensions of interfacial air voids, etc.). Nevertheless, the removal or reduction of beam hardening effects is crucial for further studies that would focus on the micro-scale details of the SCI as well as on quantitative analysis of high number of images. The results of the segmentation based on the bivariate histogram of the present 2D cross-section are visible in Figure S3.

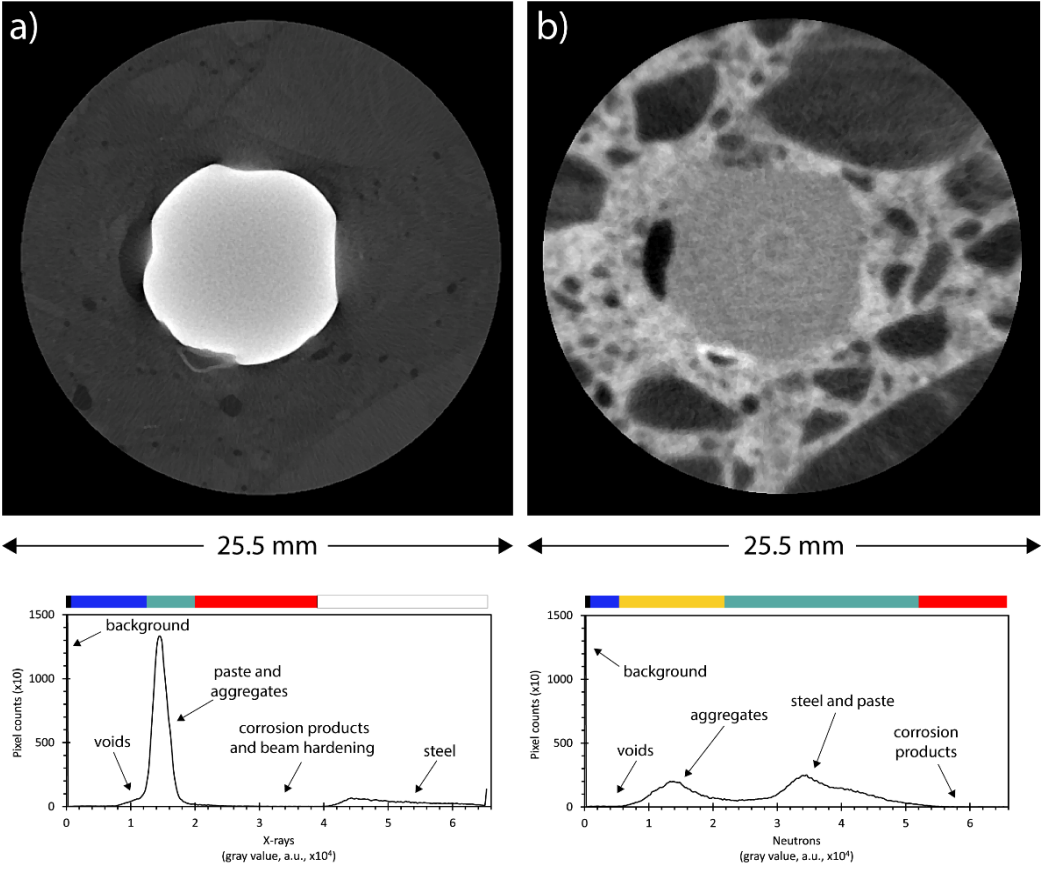

*Figure S2. 2D X-ray (a) and neutron (b) cross-section images, with respective grayscale value (GSV) histograms and color ranges of the different segmented components.*

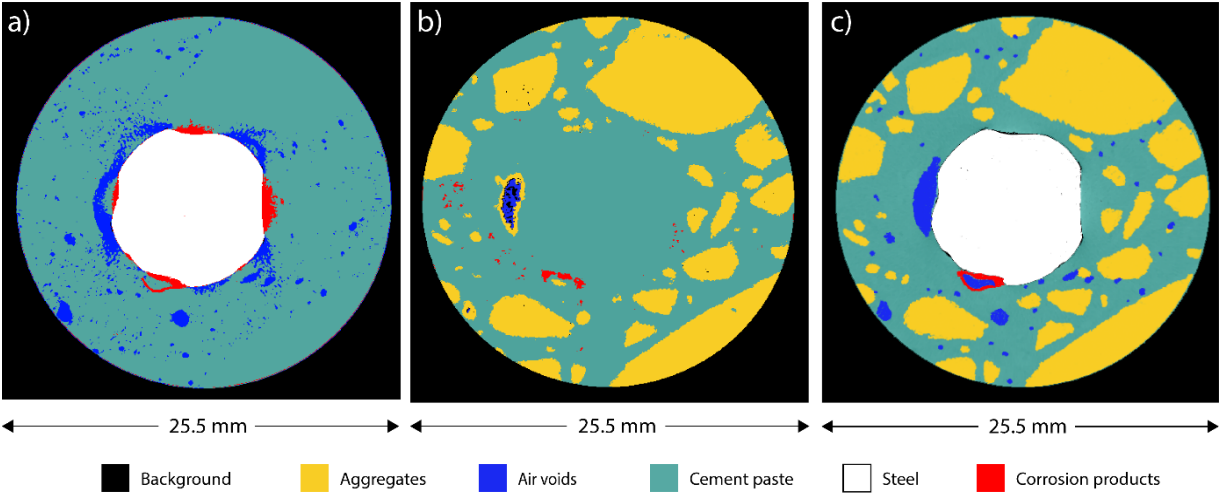

*Figure S3. Results of the segmentation procedure using the X-rays GSV histogram (a), the neutron SGV histogram (b), and the respective bivariate histogram (c).*

3